\documentclass[twocolumn,showpacs,preprintnumbers,amsmath,amssymb,superscriptaddress,prl]{revtex4}

\usepackage{graphicx}
\usepackage{dcolumn}
\usepackage{bm}
\newcommand{\ket}[1]{\mbox{$\vert #1 \rangle$}}
\newcommand{\Rb}{\mbox{$^{87}$Rb}}

\begin{document}


\title{Entanglement interferometry for precision measurement of atomic scattering properties}
\author{Artur Widera}\email[Electronic addresses: ]{Artur.Widera@Physik.Uni-Muenchen.DE; http://www.mpq.mpg.de/~haensch/bec}
\author{Olaf Mandel}
\affiliation{Ludwig-Maximilians-Universit\"at, Schellingstrasse
4/III, 80799 Munich, Germany} %
\affiliation{Max-Planck-Institut f\"ur Quantenoptik, 85748
Garching, Germany}
\author{Markus Greiner}
\affiliation{JILA, University of Colorado, Boulder, CO 80309-0440,
USA}
\author{Susanne Kreim}
\author{Theodor W.\ H\"ansch}
\affiliation{Ludwig-Maximilians-Universit\"at, Schellingstrasse
4/III, 80799 Munich, Germany} %
\affiliation{Max-Planck-Institut f\"ur Quantenoptik, 85748
Garching, Germany}
\author{Immanuel Bloch}
\affiliation{Ludwig-Maximilians-Universit\"at, Schellingstrasse
4/III, 80799 Munich, Germany}%
\affiliation{Max-Planck-Institut f\"ur Quantenoptik, 85748
Garching, Germany} %
\affiliation{Johannes-Gutenberg-Universit\"at, Staudingerweg 7,
55128 Mainz, Germany}
\date{\today}

\begin{abstract}
We report on a two-particle matter wave interferometer realized
with pairs of trapped \Rb\ atoms. Each pair of atoms is confined
at a single site of an optical lattice potential. The
interferometer is realized by first creating a coherent
spin-mixture of the two atoms and then tuning the inter-state
scattering length via a Feshbach resonance. The selective change
of the inter-state scattering length leads to an entanglement
dynamics of the two-particle state that can be detected in a
Ramsey interference experiment. This entanglement dynamics is
employed for a precision measurement of atomic interaction
parameters. Furthermore, the interferometer allows to separate
lattice sites with one or two atoms in a non-destructive way.

\end{abstract}

\pacs{03.75.Gg, 03.75.Lm, 03.75.Mn, 34.50.-s}   
\maketitle

The controlled creation of entanglement is one of the most subtle
and challenging tasks in modern quantum mechanics, with both wide
reaching practical and fundamental implications. In neutral atom
based systems, significant progress has been made during recent
years in the generation of large spin-squeezed samples of atomic
gases \cite{Julsgaard01} or the controlled creation of
Greenberger-Horne-Zeilinger (GHZ) states \cite{Greenberger90} in
cavity QED systems \cite{Rauschenbeutel00}. In addition, it has
been recognized early on that in binary spin-mixtures of
Bose-Einstein condensed quantum gases (with spin states
$|0\rangle$ and $|1\rangle$) a large amount of entanglement could
be created by controlling the difference in interaction strengths
$\chi = 1/2\,(U_{00} + U_{11} - 2 U_{01})$ between the particles
in different spin states
\cite{Sorenson01a,Jaksch99,Helmerson01,You03,Sorenson99}. Here
$U_{ij}$ denotes the interaction matrix element between atoms in
spin states $i$ and $j$. Such control can either be achieved by
moving atoms on different sites in spin-dependent optical lattice
potentials \cite{Jaksch99,Sorenson99,Brennen99,Mandel03a,Mandel03}
 or by tuning the scattering lengths, such that $\chi\neq0$ \cite{Sorenson01a,You03,Micheli03}.
In the latter case, the simple creation of a coherent
spin-mixture, e.g. by an initial $\pi/2$-pulse, followed by a
subsequent evolution of the spin system, would automatically lead
to highly spin-squeezed or entangled $N$-Particle GHZ-like states.
Here we demonstrate such entanglement dynamics with pairs of atoms
trapped in the ground state of a potential well in an optical
lattice. Such pairs form a unique and highly controllable model
system to study interactions between two particles. By using a
recently predicted inter-state Feshbach resonance in $^{87}$Rb
\cite{Kempen02} we are able to control $\chi$ and observe the
ensuing entanglement evolution in a Ramsey type experiment. We
show that this dynamical evolution of the atom pairs into
entangled and disentangled states can be used to obtain precise
information on the scattering properties of the systems. The
entanglement interferometer makes it furthermore possible to
separate singly occupied lattice sites from doubly occupied sites
in a non-destructive way.

\begin{figure}
\includegraphics{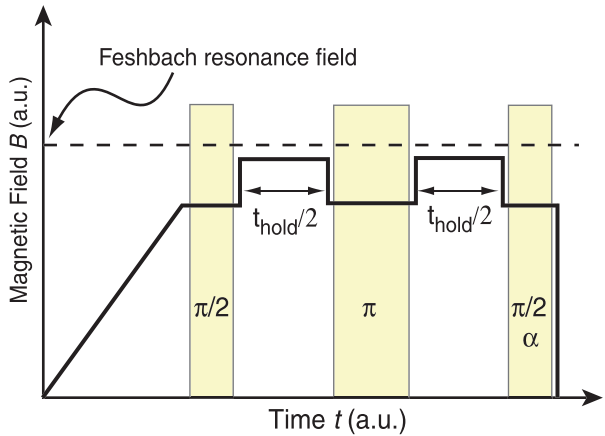}
\caption{\label{fig:Sequence} Ramsey interferometer sequence
(shaded bars). Ramsey fringes are obtained by varying the phase
$\alpha$ of the last microwave $\pi/2$-pulse. The magnetic field
can be ramped to different magnetic field values in order to
change the interaction properties of the atoms.}
\end{figure}

The Ramsey interferometer sequence used in the experiment consists
of two $\pi/2$-pulses with pulse separation $t_{\text{hold}}$,
where the phase $\alpha$ of the last pulse can be varied. Between
the two $\pi/2$-pulses the interaction behavior $\chi$ of the
atoms is modified by using a Feshbach resonance occurring between
atoms in hyperfine states $\ket{0} \equiv \ket{F=1, m_F=+1}$ and
$\ket{1} \equiv \ket{F=2, m_F=-1}$ \cite{Kempen02,Erhard03}. Here
$F$ and $m_F$ denote the total angular momentum and its
projection, respectively. In the case of a single isolated atom,
initially in the state $\psi^{i}_1=\ket{0}$, the effect of a
changing interaction behavior has no consequence for the particle
since it does not interact with other particles. After applying
the experimental sequence shown in Fig.\,\ref{fig:Sequence} the
probability of finding the atom in state \ket{1} is simply given
by $P ( \psi^f_1, \alpha ) = \frac{1}{2} \left( 1 - \cos \alpha
\right)$, which describes the usual Ramsey fringe without
decoherence.
In the case of two particles, the change of interaction parameters
leads to an additional entangling-disentangling dynamics which is
markedly different. Let us consider two atoms in the internal
state $\psi^i_2=\ket{0} \otimes \ket{0} \equiv \ket{00}$. First a
$\pi/2$-pulse places the atoms in a coherent superposition of the
two internal states $\psi_2 = \frac{1}{2} \left( \ket{00}
-\ket{01} - \ket{10} + \ket{11} \right)$. After a time $t$ each
two-particle state obtains a phase factor $e^{-i \phi_{ij}}$ due
to interactions, where $\phi_{ij} = \frac{U_{ij}}{\hbar}\cdot t$
is the collisional phase shift, with $U_{ij} = (4\pi \, \hbar^2 \;
a_{ij})/m \times \int d^3 x\; \vert \psi_i \vert^2 \vert \psi_j
\vert^2 $ being the on-site interaction matrix element. Here
$a_{ij}$ represents the elastic scattering length between
particles in states $i$ and $j$, $\psi_{i(j)}$ is the ground state
wave function of an atom in spin state $i(j)$ and $m$ the mass of
a single atom. For $t=t_{\text{hold}}/2$ the two particle state
then evolves into $\psi_2 = \frac{1}{2} (e ^{-i
\frac{\phi_{00}}{2}} \ket{00} - e ^{-i \frac{\phi_{01}}{2}}
\ket{01} - e ^{-i \frac{\phi_{01}}{2}} \ket{10} + e ^{-i
\frac{\phi_{11}}{2}} \ket{11})$, where $\phi_{01} = \phi_{10}$.
After a spin echo $\pi$-pulse, a further interaction time
$t_{\text{hold}}/2$ and a last $\pi/2$-pulse the final state reads
$\psi_2^f = \frac{1}{2} \left \{ c_c^+ \ket{00} - c_s (\ket{01} +
\ket{10}) + c_c^- \ket{11} \right\}$, where $c_c^{\pm} \equiv
e^{\mp i \alpha} (\cos \alpha \pm e^{-i \phi_\chi})$, $c_s \equiv
i \sin \alpha$ and $\phi_\chi \equiv -(\phi_{00} + \phi_{11} - 2\,
\phi_{10})/2$. The probability of finding an atom in state \ket{1}
can then be expressed by $P ( \psi^f_2, \alpha, \phi_\chi ) =
\frac{1}{2} ( 1 - \cos \alpha \cdot \cos \phi_\chi)$, which is
modulated in amplitude compared to the case of single atoms.

Four main cases illustrate the dynamics of the two-particle
system: {\bf (i)} For $\phi_\chi = 0$ the Ramsey fringe $P (
\psi^f_2, \alpha, \phi_\chi = 0 ) = \frac{1}{2} \left( 1 - \cos
\alpha \right)$ is identical to the fringe of an isolated particle
as shown above. {\bf (ii)} If the interactions lead to a phase
difference of $\phi_\chi = \pi/2$, the final state $\psi^f_2(
\phi_\chi = \pi/2)$ is a maximally entangled Bell-like state. For
such a state the corresponding Ramsey fringe $P ( \psi^f_2,
\alpha, \phi_\chi = \pi/2 ) = \frac{1}{2}$ does not exhibit any
modulation \cite{Mandel03}. {\bf (iii)} When the phase difference
is increased to $\phi_\chi = \pi$, the system is disentangled
again, and the corresponding Ramsey fringe $P ( \psi^f_2, \alpha,
\phi_\chi = \pi ) = \frac{1}{2} \left( 1 + \cos \alpha \right)$ is
phase shifted by $\pi$ with respect to the case of a single
particle. It should be noted that for this interaction phase the
state vectors of isolated atoms and interacting atom pairs are
orthogonal to each other. Therefore, by choosing a specific single
particle phase, either species can be transferred into the
\ket{F=2} state and removed by a subsequent resonant laser pulse.
The remaining atoms would form a pure lattice of either single
atoms or atom pairs which again can evolve to Bell-like pairs. For
even larger phase differences, the system entangles and
disentangles again, until for {\bf (iv)} $\phi_\chi = 2\pi$ the
system exhibits a Ramsey fringe which is in phase with the fringe
of a single atom. In a system containing $N_1$ isolated single
atoms and $N_2/2$ isolated pairs of atoms, the total fringe will
be a superposition of the two distinct fringes and will have a
visibility according to:

\begin{equation}\label{eq:Vis}
V(\chi) = V_0 \cdot e^{-\frac{t}{\tau_1}} \left \{ (1-n_2) + n_2
\cdot e^{-\frac{t}{\tau_2}} \; \cos \phi_\chi \right \},
\end{equation}

where we have included decoherence and two-body losses with time
constants $\tau_1$ and $\tau_2$, respectively, $V_0$ is a finite
initial visibility \footnote{Here the visibility is defined as
$V=(P^{tot}_{max}-P^{tot}_{min})/(P^{tot}_{max}+P^{tot}_{min})$.}
and $n_i = N_i/(N_1+N_2)$, $i=1,2$ with $n_1 + n_2 =1$. Whereas
the contribution of isolated atoms to the fringe visibility $V$
remains unaffected under a change of $\chi$, the total fringe
signal shows a dynamics with the same periodicity as
$P(\psi_2^f)$. For zero phase difference $\phi_\chi=0$ we expect
to measure a Ramsey fringe with high visibility $V$. The
visibility decreases for increasing $\phi_\chi$ and reaches a
minimum for $\phi_\chi= \pi$ where the fringes from single atoms
and from atom pairs are out of phase and partially compensate each
other in the total signal. For larger interaction phase
differences, the total visibility increases, until it shows a
maximum for $\phi_\chi = 2 \pi$, where the two Ramsey fringes are
completely in phase again. It should be noted that the interaction
time $t_R$ after which $\phi_\chi = 2 \pi$ depends on the
difference in the interaction matrix elements $U_{ij}$, and for
constant overlap of the wave functions on the elastic scattering
length difference $\Delta a_{s,\chi} \equiv
\frac{1}{2}(a_{00}+a_{11}-2 a_{10})$.\\

The experimental setup is similar to our previous work
\cite{Greiner02a,Greiner02c}. We start with a BEC of up to $3
\times 10^{5}$ \Rb\ atoms trapped in the hyperfine ground state
\ket{F=1, m_F=-1}. We load the BEC into a pure two or three
dimensional optical lattice potential formed by three mutually
orthogonal standing waves of far detuned light. The wavelengths
used for the different standing waves are 829\,nm ($y$ and $z$
axes) and 853\,nm ($x$ direction), with trapping frequencies at
each lattice site of $\omega_x =2\pi\times 33$\,kHz, $\omega_y
=2\pi\times43$\,kHz, and $\omega_z =2\pi\times41$\,kHz. In order
to preserve spin polarization of the atoms in the optical trap, we
maintain a 1\,G magnetic offset field along the $x$-direction. The
atoms are prepared in the Feshbach resonance sensitive spin
mixture by transferring the population from the \ket{F=1, m_F=-1}
state into the $\ket{0} \equiv \ket{F=1, m_F=+1}$ hyperfine level
via a radio frequency (RF) Landau-Zener sweep. We then increase
the magnetic field to 8.63\,G. By applying a microwave field
around 6.8 GHz and RF radiation around 6 MHz, we are able to
coherently couple the two internal states $|0\rangle$ and
$|1\rangle$ with a two-photon transition similar to
\cite{Hall98a}.

\begin{figure}
\includegraphics{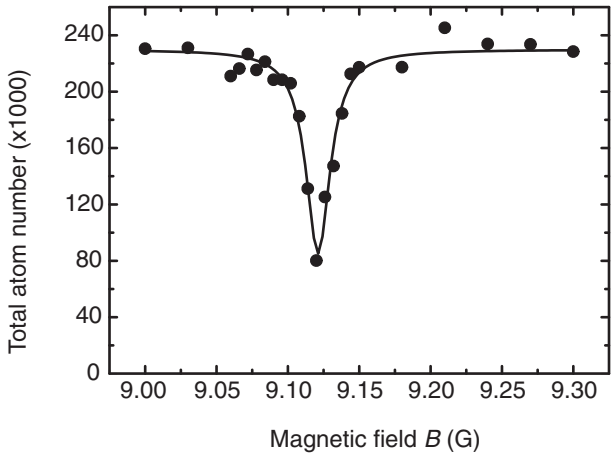}
\caption{\label{fig:Verlust} Measurement of total atom number
versus magnetic field in a two dimensional optical lattice. The
solid line is a lorentzian fit to the data with center at
9.121(5)\,G and a width of 20(5)\,mG. The hold time at the various
magnetic field values is 1\,ms.}
\end{figure}

\begin{figure}
\includegraphics{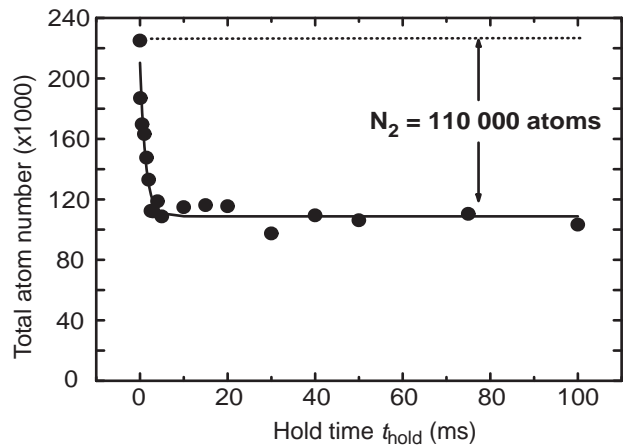}
\caption{\label{fig:Gitter3D} Time resolved measurement of total
atom number at the measured Feshbach resonance in a three
dimensional lattice potential. Sites with more than one atom are
emptied within 3\,ms, whereas sites with only one atom are
protected from loss. The solid line is a fit to an exponential
decay with offset from which we find to have a ratio
$N_1/N_2\approx 1.1$.}
\end{figure}

In order to locate the position of the Feshbach resonance through
enhanced atomic losses, we load the BEC into a two dimensional
optical lattice potential. The atomic density is thereby strongly
increased compared to a simple dipole trap, thus loss processes
occur with higher probability. The magnetic field is subsequently
increased to different values within 10\,$\mu$s. After holding the
atoms for 1\,ms at a specific magnetic field, we switch off all
trapping potentials and magnetic fields to measure the remaining
total atom number in a time-of-flight (TOF) measurement (see
Fig.\,\ref{fig:Verlust}). The magnetic field has been calibrated
by measuring the frequency of the $\ket{F=1, m_F=-1} \rightarrow
\ket{F=2, m_F=-2}$ microwave transition at different magnetic
field values and employing the Breit-Rabi formula to determine the
actual field strength. Due to background magnetic field
fluctuations, the magnetic field calibration has an uncertainty of
3\,mG, and noise of the magnetic field creating current source
introduces an additional uncertainty of 2\,mG. The measured
position of 9.121(5)\,G  of the resonance agrees well with the
predicted value of 9.123\,G within our measurement uncertainty
\footnote{The inter-state Feshbach resonance has recently also
been detected through atom loss measurements by Erhard et al.
\cite{Erhard03}. There the resonance was found to be located at a
magnetic field of 9.08(1)\,G, which disagrees with our result.}.
In order to determine the ratio of single to paired atoms in our
three dimensional lattice potential, we monitor the loss of atoms
when we hold the atomic sample at the resonance magnetic field for
a variable time (see Fig.\,\ref{fig:Gitter3D}).
Lattice sites which are occupied by more than one atom are
depleted within 3 ms due to the increased two- and three-body
collision rates and the high density at single lattice sites.
Isolated atoms, however, are protected from collisions and remain
trapped. Assuming a negligible number of sites with three atoms, a
fit with an exponential decay yields a time constant of
$1.3(2)$\,ms and a ratio $N_1/N_2\approx 1.1$.

\begin{figure}
\includegraphics{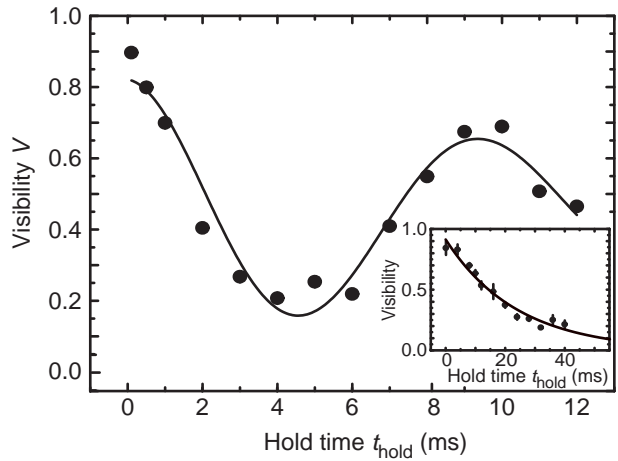}
\caption{\label{fig:Visplot} Measured visibility dynamics due to
entanglement in the system for a magnetic field value of 9.081 G.
The solid line is a fit using eq.~(\ref{eq:Vis}) that yields a
revival time $t_R$ of $9.5(2)$\,ms which depends on interaction
properties. The reduced visibility is due to decoherence. The
inset is a measurement of the Ramsey sequence for a constant
magnetic field $B= 8.63$\,G, away from the Feshbach resonance and
the solid line is a fit describing an exponential decay of the
fringe visibility.}
\end{figure}

In order to determine the elastic scattering properties we apply
the Ramsey interferometer sequence that has been described earlier
(see Fig.\,\ref{fig:Sequence}). At each magnetic field value we
record the Ramsey fringe visibility for different interaction
times $t_{\text{hold}}$ (see e.g.~Fig.\,\ref{fig:Visplot} for a
field of $B$=9.081\,G).
The revival time $t_R$ at which the visibility shows its maximum
is determined by a fit using eq.~\ref{eq:Vis}. This revival time
depends on the difference in the interaction matrix element. A
single revival of the fringe visibility could in principle also be
caused by a complete loss of lattice sites with two atoms. We have
however checked that even for the more pronounced losses at the
Feshbach resonance the system exhibits dynamics due to interaction
after the first revival. In this case losses would shift the
revival time by less than 1\%.

\begin{figure}
\includegraphics{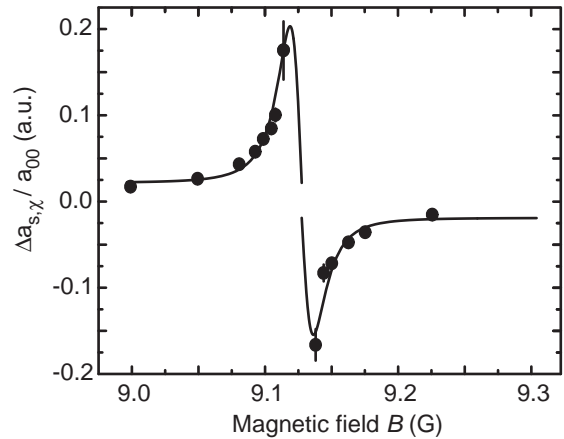}
\caption{\label{fig:Elastisch} Measured change of elastic
scattering length for magnetic field values around the Feshbach
resonance. Both sides were fitted separately to the expected
behavior with a common center (9.128(5)\,G) and width
(15(4)\,mG).}
\end{figure}

In order to extract information on the changes in the scattering
length from the revival times, one can measure the the on-site
matrix element $U_{00}$ through a collapse and revival experiment
that we have demonstrated earlier \cite{Greiner02c}. For the same
experimental parameters, we find $U_{00}=h/396(11)\,\mu$s. Using
this information we can calculate $\chi/U_{00}=\Delta
a_{s,\chi}/a_{00}$, which expresses the change in scattering
length measured in units of the scattering length $a_{00}$. In
order to map out the change in the elastic scattering length on
the Fesh\-bach resonance, $\Delta a_{s,\chi}/a_{00}$ has been
measured for several magnetic fields and is shown in
Fig.\,\ref{fig:Elastisch}. Since the entanglement interferometer
can only measure absolute values of $\Delta a_{s,\chi}/a_{00}$, we
perform a usual time of flight measurement to obtain information
on the sign of the scattering length differences. For this, we
leave the Feshbach field switched on for the first 3\,ms of the
TOF \cite{Inouye98}. During this time the altered interaction
energy is converted into kinetic energy, and the size of the atom
cloud is measured. For magnetic fields below the Feshbach
resonance, we have found the cloud to be slightly larger (9\%
change in size) in the axial direction, whereas it is slightly
smaller above the Feshbach resonance. From this we conclude that
the interspecies scattering length grows and shrinks below and
above the Feshbach resonance field respectively.

Since we ramp the magnetic field through the resonance in order to
address the region above the resonance, we fit both branches of
the scattering length change separately to a dispersive profile,
with the constraint of a common center and width. The fit yields a
center of 9.128(5)\,G and a width of 15(4)\,mG, in good agreement
with our loss measurements. The two branches show an offset from
the $\chi=0$ line which arises from the non-zero scattering length
difference $\delta a_s \equiv |a_{00} - a_{11}|$, which in
principle can be extracted with high precision if decoherence
permitted for longer interaction times.

The interferometric method presented allows for high precision
measurements of relative changes of the scattering lengths. In
order to demonstrate this, we assume an error-free scattering
length $a_{00} = 100.4\,a_0$ \footnote{S.~Kokkelmans, private
communication.}, where $a_0$ is the Bohr radius. With this, we
determine the change of elastic scattering length for $B=$
9.081\,G to be $\Delta a_{s,\chi}=4.2(1)\,a_0$.

In conclusion, we have presented a novel interferometric method to
create and investigate entanglement dynamics in an array of spin
mixtures of neutral atoms. The observed entanglement oscillations
allow the precise determination of interaction properties between
atoms in different spin states. We have demonstrated the
versatility of the interferometer by characterizing the elastic
scattering properties of a newly predicted weak inter-state
Feshbach resonance in \Rb. We have found both the elastic and
inelastic channel of the measured Feshbach resonance to be in good
agreement with the theoretical prediction. The two-particle
interferometer furthermore enables the direct creation of arrays
of Bell-states together with the non-destructive separation of
singly from doubly occupied sites.

We would like to thank Anton Scheich for support with the
electronics and Tim Rom and Thorsten Best for assistance with the
experiment. We also acknowledge financial support from the
Bayerische Forschungsstiftung and AFOSR.

\end{document}